\documentclass[twocolumn,showpacs,preprintnumbers,amsmath,amssymb]{revtex4}
\usepackage[dvips]{graphicx}
\usepackage{dcolumn}

\begin{document}

\title
{Modulated vortex states in Rashba non-centrosymmetric superconductors
}

\author{Yuichi Matsunaga, Norihito Hiasa, and Ryusuke Ikeda}

\affiliation{%
Department of Physics, Kyoto University, Kyoto 606-8502, Japan
}

\date{\today}

\begin{abstract} 
Vortex lattice structures to occur in Rashba non-centrosymmetric superconductors under a magnetic field {\it parallel} to the basal plane are studied by assuming the $s$-wave pairing and taking account of both the paramagnetic and orbital depairings. A high field vortex lattice with a modulation perpendicular to the field and of Fulde-Ferrell-Larkin-Ovchinnikov type, occurring in the limit of vanishing ${\tilde \zeta}$ (spin-orbit coupling {\it normalized} by Fermi energy), is found to be replaced with increasing ${\tilde \zeta}$ by another new vortex lattice with a modulation of the gap {\it amplitude} and induced by the absence of inversion symmetry. A correlation between a structural transition and an inflection of the $H_{c2}(T)$-curve is discussed in relation to $H_{c2}$ data of CeRhSi$_3$. 
\end{abstract}



\maketitle

Recently, superconductivity in materials with no inversion symmetry has been a focus of intense interest \cite{GR,Bauer}. A spin-orbit coupling due to the absence of an inversion symmetry may lead to coexistence of spin-singlet and triplet pairing states and result in new phenomena stemming from the pairing symmetries \cite{Yanase,Fujimoto}. Another interest is a consequence of the non-centrosymmetry in superconductivity in nonzero magnetic fields. An interplay of the Zeeman effect and a splitting of Fermi surface (FS) due to the spin-orbit coupling was shown to, in fields (${\bf H} \perp c$) parallel to the basal ($a$-$b$) plane, lead to a helical phase modulation \cite{Kaur} of the superconducting order parameter $\Delta({\bf r})$ similar to the Fulde-Ferrell state \cite{FF} in a Pauli-limited superconductor as far as the densities of states $N_j$ on the splitted two FSs, denoted by $j=1$, $2$, are different from each other. Later, the mixed state was studied by invoking the Pauli limit \cite{Agter} with {\it no} vortices, and, just as in the centrosymmetric case, the presence of a modulation of the amplitude $|{\Delta}|$ of Larkin-Ovchinnikov (LO) type \cite{LO} was argued which is separated by a {\it single} transition from the phase-modulated state in lower fields. However, it is natural to expect appearance of a different superconducting phase from those in the centrosymmetric case. 

In this work, mixed or vortex states occurring in Rashba non-centrosymmetric materials \cite{GR} in ${\bf H} \perp c$ are investigated in clean limit by assuming the $s$-wave pairing and incorporating both the paramagnetic and orbital depairings. Based on existing estimates \cite{Fujimoto} of the spin-orbit coupling energy $\zeta$ in real non-centrosymmetric materials, the relation $E_{\rm F} \gg \zeta \gg T_c$ between relevant energy scales will be reasonably assumed, where $E_{\rm F}$ is the Fermi energy, and $T_c$ is the zero field ($H=0$) superconducting transition temperature. Then, the key parameter measuring the broken inversion symmetry in superconductivity is $\delta N = |N_1-N_2|/(N_1+N_2)$ which is of the order of ${\tilde \zeta} \equiv \zeta/E_{\rm F}$ rather than $k_{\rm B} T_c/\zeta$. The main focus in ${\bf H} \perp c$, where the paramagnetic depairing is effective, is to understand effects of the absence of inversion symmetry on a modulated state of LO type in higher $H$ and at lower temperatures. Our findings are summarized as follows: In the specific limit, $\delta N \to 0$, the LO-like state with a modulation perpendicular to ${\bf H}$ occurs as a result of the paramagnetic depairing and is transformed with decreasing $H$ into an Abrikosov lattice through two first order structural transitions (FOSTs). However, changes of the lattice structure obeying the shift of FSs induced by the Zeeman energy do not occur. Once a small but nonzero $\delta N$ is taken into account, however, the LO-like state is pushed up to higher $H$, and a new intermediate phase appears accompanied by an amplitude modulation parallel to the direction of the helical phase modulation as a consequence of the lack of inversion symmetry. As $|\delta N|$ is increased so that the LO-like state disappears, the high field side of the $H$-$T$ phase diagram is entirely occupied by this new modulated state. By comparing these results with those in the Pauli limit \cite{Agter}, essential roles of the orbital depairing in this issue are emphasized. 

We start from a BCS-like Hamiltonian 
\begin{eqnarray}
{\cal H}_{\rm BCS} &=& \! \sum_{{\bf k}, s_1, s_2} c_{{\bf k}, s_1}^\dagger (  \varepsilon_{\bf k} \, \delta_{s_1, s_2} \! + \! (\zeta \, {\hat {\bf g}}_{\bf k} + \mu_B {\bf H})\cdot{\bf \sigma}_{s_1, s_2} ) \, c_{{\bf k}, s_2} \nonumber \\
&-& |\lambda| \sum_{\bf q} \, \Psi^\dagger_{\bf q} \, \Psi_{\bf q}
\end{eqnarray}
where $\mu_B H$ is the Zeeman energy, $\varepsilon_{\bf k}$ denotes the {\it bare} band energy, and ${\hat {\bf g}}_{\bf k}$ is the unit vector $({\hat {\bf k}} \times {\hat z})$ with ${\hat z}$ perpendicular to the basal ($a$-$b$ or $x$-$y$) plane. A Ginzburg-Landau (GL) free energy $F$ will be derived within the mean field approximation and in type II limit with no variation of flux density. As performed elsewhere \cite{GR}, diagonalization of the quadratic term leads to a splitting of FS into two branches ($j=1$, $2$), 
and, consistently with this, the pair-field operator $\Psi_{\bf q} = -{\rm i} \sum_{\bf k} (\sigma_y)_{\beta \alpha} c_{-{\bf k} \, + \, {\bf q}/2 \, \beta} \, c_{{\bf k} \, + \, {\bf q}/2 \, \alpha}$ may be rewritten in terms of the field operator $a_{{\bf k} \, j}$ on each band as $\sum_{\bf k} \sum_j f_j({\bf k}) a_{-{\bf k} \, + \, {\bf q}/2 \, j} \, a_{{\bf k} \, + \, {\bf q}/2 \, j}$ \cite{Feigel}, where $f_j({\bf k})$ satisfies $|f_j|=1$, and its detailed expression is not necessary within the weak-coupling BCS theory. In reaching the latter expression, a small contribution due to interband interactions of O($\varepsilon_H^3$) to $F$ was neglected, where $\varepsilon_H = {\rm Max}(\mu_B H, T_c)/\zeta$. By replacing the interaction term in eq.(1) by $\int d^3r ( |\lambda|^{-1} |\Delta|^2 - \Psi^\dagger \Delta - \Psi \Delta^* )$ in terms of the order parameter $\Delta$, expanding the free energy in $\Delta$ results in $F$. Its quadratic term is given by 
\begin{eqnarray}
F_2 &=& \! \! \int d^3r \Delta^*({\bf r}) \biggl[ \frac{1}{|\lambda|} - \int_0^\infty \! \! d\rho \frac{2 \pi T N \langle {\cal O}(\rho; {\bf \Pi}_Q) \rangle}{{\rm sinh}(2 \pi \rho T)} \biggr] \Delta({\bf r}), \nonumber
\end{eqnarray}
where $N=N_1+N_2$, $\langle \,\,\, \rangle$ denotes the average over the momentum ${\bf k}$ on FSs, and 
\begin{eqnarray}
{\cal O}(\rho; {\bf \Pi}_Q) &=& \frac{1}{2} \! 
\sum_{\sigma=\pm1} \Biggl[ \exp(-{\rm i}\rho \sigma v_x Q) (\, {\rm cos}(2 \mu_B H {\hat k}_x \rho) \\
&+& {\rm i} \sigma \, \delta N \, {\rm sin}(2 \mu_B H {\hat k}_x \rho) \, )  \Biggr] \exp(-{\rm i} \sigma \rho{\bf v}\cdot{\bf \Pi}_Q) \nonumber
\end{eqnarray}
under ${\bf H}$ parallel to ${\hat y}$, where ${\bf \Pi}_Q = - {\rm i}\nabla + 2 e {\bf A} - Q {\hat x}$, ${\bf v}={\bf v}_\perp + v_z {\hat z}$ is the velocity vector on FSs, and a possible modulation wave vector ${\bf Q} \equiv Q {\hat x}$ was introduced (see below). The ${\hat k}_x$-dependence accompanying $\mu_B H$ implies that the paramagnetic depairing does not work effectively in the free energy terms associated with the gradient parallel to ${\bf H}$. Thus, in contrast to the centrosymmetric case \cite{I1,AI}, $\Delta({\bf r})$ solutions modulating along ${\bf H}$ {\it never} appear in Rashba superconductors. In $H \ll H_P(0)$, where $H_P(0) = \Delta(0)/\sqrt{2} \mu_B$ is the Pauli-limiting field, the factor prior to $\exp(-{\rm i}\sigma \rho{\bf v}\cdot{\bf \Pi}_Q)$ is approximated by unity if $Q= Q_0 \equiv 2 \mu_B H \, \delta N/|{\bf v}_\perp|$ \cite{Kaur}, and then, $\Delta$ is diagonalized in terms of Landau levels (LLs) with even indices, $\Delta = \sum_{n} a_{2n} \, \phi_{2n}({\bf r} - Q_0 r_H^2{\hat z}|0)$, where $r_H^{-2}=2e H$, and $\phi_0({\bf r}|0)$ is an Abrikosov lattice solution in the lowest ($n=0$) LL and is one of magnetic Bloch states \cite{I1}, $\phi_n({\bf r}|{\bf r}_0)$, in the same gauge ${\bf A}=H z{\hat x}$. Note that, using the property $\phi_n({\bf r}|{\bf r}_0) = \exp({\rm i}z_0 x/r_H^2) \, \phi_n({\bf r}+{\bf r}_0|0)$ \cite{I1}, we have the form of the helical state \cite{Kaur}, $\Delta = \exp({\rm i} Q_0 x) \, \sum_n a_{2n} \phi_{2n}({\bf r}|-Q_0 r_H^2{\hat z})$. Therefore, the helical {\it phase} modulation \cite{Kaur} in $H \ll H_P$ merely corresponds to a {\it uniform} displacement 
of the vortices. 
By contrast, in $H > H_P(0)$, the $\delta N$ dependent imaginary terms cannot be erased from the factor prior to $\exp(-{\rm i}\sigma\rho{\bf v}\cdot{\bf \Pi}_Q)$ in eq.(2) any longer and result in a $\Delta$-solution represented by both of even {\it and odd} LLs. This hybridization of even and odd LLs in equilibrium is a new feature stemming from the lack of inversion symmetry and, as is shown below, leads to a new vortex state. 

Hereafter, we use a bare FS of the form \cite{I1} of a corrugated cylinder extending along ${\hat z}$ and with a coherence length anisotropy $\gamma=\xi_x/\xi_z = 6$, where $\xi_\mu$ is the coherence length in $\mu$-direction. However, no layer structure is assumed in real space. 
Quantitatively similar results should follow in terms of an ellipsoidal FS with a similar $\gamma$-value as far as ${\bf H} \perp {\hat z}$. The paramagnetic depairing is measured by the Maki parameter in ${\bf H} \parallel {\hat z}$, $\alpha_{\rm M}^{(z)} \simeq 7.1 \mu_B H_{\rm orb}^{(2D)}(0)/(2 \pi T_c)$, where $H_{\rm orb}^{({\rm 2D})}(0)$ is the 2D orbital-limiting field. Note that the argument of the trigonometric functions in eq.(2) is proportional to $\alpha_{\rm M}^{(z)} \, h$, where $h=H/H_{\rm orb}^{({\rm 2D})}(0)$. At least in $\alpha_{\rm M}^{(z)} > 2$, the high field state in low $T$ is, as in various centrosymmetric systems \cite{Klein,AI}, described by {\it higher} LLs. 

In examining the details of $H$-$T$ phase diagrams by including higher LLs, the best approach will be to diagonalize $F_2$ in terms of $\Delta=\sum_n a_n \phi_n({\bf r}-Q r_H^2 {\hat z}|0)$ with the consistent minimization in $Q$, $\partial F/\partial Q=0$ \cite{Feigel} implying that the total current is zero, where $\phi_0({\bf r}|0) = \sum_p \exp[ \, {\rm i}kpx - \gamma( \! z + k p \, r_H^2 \!)^2/(2 r_H^2) + {\rm i}(k^2 r_H^2 p^2 {\rm cot}\theta/2) \, ]$, and the parameters $k$ and $\theta$ determine the vortex lattice structure. Due to inclusion of numerous LLs, however, it is not tractable to perform the $Q$-minimization exactly. As an alternative approach, we first choose $Q=Q_0$ and try to cure the deviation $Q-Q_0$ by incorporating more numerous LLs. In this procedure, a finite uniform current due to the deviation $Q-Q_0$ should be cancelled by the corresponding one \cite{I1} arising from the even-odd LL mixings \cite{HI2}. Besides, inclusion of more LLs are needed, even when $\delta N=0$, with increasing $\alpha_{\rm M}^{(z)}$ \cite{Klein,AI}. Thus, a truncation of LLs might lead to an ambiguity in numerical results with increasing $|\delta N|$ or $\alpha_{\rm M}^{(z)}$. However, we have verified in the range $\alpha_{\rm M}^{(z)} < 5$ and for $0 \leq \delta N \leq 0.2$ that vortex lattice structures following from eight LLs ($n \leq 7$) are essentially unaffected by adding higher two LLs further. Below, we will focus on $\Delta$ expressed by the eight LLs with $Q=Q_0$. Then, by diagonalizing $F_2$, the resulting superposition of LLs with the lowest energy is chosen at each ($T$,$H$) as a candidate of $\Delta$ in equilibrium. We note that the details of FSs are encoded as the relative weight between LLs in $\Delta$. 

Next, after substituting the resulting candidate of $\Delta$ into the quartic term $F_4$ of $F$, the parameters ($k$,$\theta$) in $\phi_0({\bf r}|0)$ are determined by minimizing $F_4$ at each ($T$,$H$). However, a full evaluation of $F_4$ is a hard task because of spatially nonlocal corrections \cite{AI} in $F_4$ and of the inclusion of numerous LLs. Fortunately, since roles of the nonlocal corrections tend to be reduced with increasing $\alpha_{\rm M}^{(z)}$ \cite{HI}, it is proper in the present case to replace $F_4$ derived microscopically by a spatially local conventional one, $F_{4,{\rm loc}} = c_4 \int d^3r |\Delta({\bf r})|^4$, where the constant $c_4$ is derived microscopically. Further, we have verified that $F_4 > 0$ by assuming $\Delta$ expressed by a single LL or by a couple of LLs as the candidate, suggesting that the $H_{c2}$-transition is of second order in Rashba superconductors. For these reasons, we have examined the lattice structures using $F_{4,{\rm loc}}$ with a positive $c_4$. Note that, to determine the stable structure at {\it each} ($T$,$H$), the $H$ and $T$ dependences of $c_4$ are not needed. 

\begin{figure}[b]
\caption{(Color online) A $h$-$t$ phase diagram for $\alpha_{\rm M}^{(z)}=2.85$ and $\delta N=0$ obtained by including LLs with $0 \leq n \leq 7$ in $\Delta$, where $t=T/T_{c0}$. Each red (blue) solid curve denotes an FOST ($H_{c2}(T)$). Insets show real space patterns of $|\Delta(x,z)|^2$ at the indicated ($t$,$h$) points. The axis of cylindrical FS is parallel to ${\hat z}$, 
while ${\bf Q} \parallel {\hat x}$. The Pauli-limiting field corresponds to $h=0.496$.} \label{fig.1}
\end{figure}

Let us start with explaining the resulting phase diagram in the limiting $\delta N=0$ case where ${\bf Q}=0$ \cite{Kaur} (Fig.1). We find that, through diagonalizing $F_2$, $H_{c2}(T)$ for $\delta N = 0$ is separated by an FOST (FOST1) into one branch above FOST1 determined by {\it only} odd LLs and another one below it due {\it only} to even LLs. We have verified that this division into the odd and even branches for $\delta N=0$ remains valid at least in the range $\alpha_{\rm M}^{(z)} \leq 7$. Above FOST1, the state below $H_{c2}$ is the LO-like phase formed by only odd LLs, while it has no additional nodal lines below FOST1 where only even LLs participate in. 
As the insets (A) and (B) show, the nodal lines in the LO-like state are removed in passing through FOST1 by decreasing $H$. In any state we obtain, the periodicity in $z$ of $|\Delta|$ appears as a periodicity of vortex positions. The energy of (A) is quite close to but slightly lower than that of a similar structure with nodal lines perpendicular to ${\bf Q}$. Namely, the orientation of nodal lines is controlled by interactions between the vortices and unaffected by the details of FSs. In fact, the spacing between neighboring nodal lines is not of O($|{\bf v}_\perp|/\mu_B H$) \cite{Agter} but of O($r_H$), reflecting the details in real space rather than in the ${\bf k}$-space. The lattice (B) is continuously transformed into a deformed square lattice (C) with a smooth rotation of the wreckage of the LO nodal lines. Figure 1 includes another weak FOST (FOST2) in lower $H$: The lattice (C) {\it discontinuously} changes through FOST2 into the familiar triangular one (D) with decreasing $H$. Since the odd LLs do not appear around FOST2, the origin of FOST2 is the mixing of even higher LLs with $n \leq 6$ becoming more important for larger $\alpha_{\rm M}^{(z)}$. This transition on the lattice symmetry might correspond to that suggested in Ref.\cite{Kaur}. However, it was not considered there \cite{Kaur,Agter} consistently with the appearance of the LO-like state in higher $H$ in the {\it same} phase diagram. Besides, the fact that some lattice structures in Fig.1 have no reflection symmetry in $z$ suggests a possibility of two lattice domains in this artificial case with $\delta N=0$. 

\begin{figure}[t]
\caption{(Color online) Corresponding ones to Fig.1 for (a) $\delta N = 0.003$ and (b) $0.2$. The state below FOST2 is the triangular lattice everywhere.} \label{fig.2}
\end{figure}

Figure 2 shows examples of the $H$-$T$ phase diagrams in $\delta N \neq 0$ 
case where the even and odd LLs are mixed with one another. Even a small enough $|\delta N|$ ($\leq 0.003$) divides the intermediate phase in Fig.1 surrounded by FOST1, FOST2, and $H_{c2}(T)$ into two regions separated by another FOST (FOST3): Once $\delta N$ becomes nonzero, a new striped phase, (B) in Fig.2 (a), begins to appear near the three transition lines surrounding (B) and (C) of Fig.1. Due to a slight increase of $|\delta N|$, the growing new phase pushes the region including (D), similar to (C) of Fig.1, down to lower temperatures (see the arrows in Fig.2 (a)), and the nodal lines of the LO-like phase (A) disappear reflecting the even-odd LL mixing. The new striped state consists of rows of vortices, separated by two neighboring stripes perpendicular to ${\bf Q}$ on which $|\Delta|$ is {\it nonvanishing}, and the inter stripe spacing is the structural period parallel to ${\bf Q}$. With decreasing $h$, this new phase shows a continuous {\it crossover} to a square lattice, (C) in (a), just above FOST2. 
Besides, a further increase of $|\delta N|$ not only expands the region of the new phase and the square lattice to shift FOST2 to lower $h$ but also elevates FOST1 up to $H_{c2}(T)$ so that the LO-like phase is lost. Since $|\delta N| \propto \zeta$, the expansion at larger $|\delta N|$ of the new phase supported by the nonzero ${\bf Q} \parallel {\hat x}$ is reasonable. Through a comparison with Fig.1, it is seen that an increase of $|\delta N|$ enhances $H_{c2}$ and compresses the structure along ${\hat z}$. Besides, Figs.2 (a) and (b) also show that, {\it irrespective of} $|\delta N|$, FOST2 merges with an inflective point of $H_{c2}(T)$ in each figure at which $d^2H_{c2}/dT^2$ changes in sign. This feature may become relevant in experimentally verifying vortex lattice structures found here. 
\begin{figure}[t]
\caption{(Color online) A $h$-$t$ phase diagram for $|\delta N|=0.003$ obtained in terms only of the $n=0$ and $1$ LLs of $\Delta$ and $\alpha_{\rm M}^{(z)} = 2.85$. } \label{fig.3}
\end{figure}

To clarify implication of the above result further, it will be compared with Fig.3, in which only the two lowest LLs ($n=0$ and $1$) are kept, and $\Delta$ is expressed by $\Delta_{01} = a_0 \phi_0 + a_1 \phi_1$. When $\delta N=0$, the situation is essentially the same as that in a centrosymmetric system near $H_{c2}$ \cite{AI}: The two LLs do not coexist in equilibrium, and a single FOST separating two states with different structural symmetries occurs. The low field state is the triangular vortex lattice with $a_1=0$, while the high field phase expressed only by the $n=1$ LL is an LO-like phase composed of an alternation of a vortex row and a nodal line \cite{Klein}. The situation is similar to the behavior occuring through FOST1 in Figs.1 and 2 (a), where the state (A), similar to this $n=1$ LL state, is dominated by the $n=1$ LL. However, it is difficult to understand the global phase diagram by concentrating on $\Delta_{01}$. In fact, as already noted, the square to triangular structure transition at FOST2 in Fig.1 is not found without including even higher LLs ($n \geq 2$), and Fig.3 shows that the new striped phase and FOST2 are not obtained within $\Delta_{01}$. 

The fact that the LO-like state with periodic nodal lines is pushed up to the high $H$ and low $T$ corner of the $H$-$T$ phase diagram to disappear with increasing $|\delta N|$ is in conflict with the corresponding result \cite{Agter} obtained with no vortices in which the LO-like phase tends to disappear {\it away} from the $H_{c2}$ line. In fact, the results in Fig.2 suggest that diappearance of the LO-like state is facilitated by the growth of the new {\it vortex} phase, absent in the Pauli limit \cite{Agter}, with an amplitude modulation parallel to ${\bf Q}$, implying that inclusion of vortices is indispensable in the present issue. On the other hand, the $H_{c2}(T)$ curve is similar to the result in the Pauli limit except the fact that it is not divergent but flattened in $T \to 0$ limit. 

Although we have neglected a $p$-wave pairing component \cite{koga} to be {\it induced} by the $s$-wave pairing due to the nonzero ${\tilde \zeta}$, the present results are still valid if the $p$-wave pairing interaction is repulsive. The same thing holds in the case dominated by the $p$-wave. In the case, presumably applicable to CePt$_3$Si, where both the singlet and triplet pairing channels are attractive and equally important, however, {\it another} even-odd LL hybridization due to a {\it field-induced} coupling between these two pairing states dominates over the hybridization appeared in eq.(2) \cite{HI1} and may change the pictures mentioned above. Results of other extensions will be reported elsewhere. 

In conclusion, the vortex lattice structures and $H$-$T$ phase diagrams to occur in Rashba non-centrosymmetric superconductors in ${\bf H} \perp c$ have been examined, and even a small value of {\it normalized} spin-orbit coupling ${\tilde \zeta}$ is found to destroy the LO-like state and replace it by a new striped vortex lattice. The obtained results, including the square to triangular lattice transition close to the inflective point of $H_{c2}(T)$, are visible in materials with standard values of the Maki parameter and anisotropy. The present results are relevant to CeRhSi$_3$ in ${\bf H} \perp c$ where the inflective $H_{c2}(T)$-curves have been found \cite{Kimura}. 

We are grateful to S. Fujimoto and Y. Yanase for discussions and to T. Saiki for supplementary calculations.


\begin{thebibliography}{99}
\bibitem{HI2} While the manuscript is being reviewed, we have verified through a direct $Q$-minimization that this argument is correct. Details will be reported elsewhere.
\end{thebibliography}
\end{document}